\documentclass[5p,final,twocolumn]{elsarticle}
\usepackage{amssymb}
\usepackage{epsfig}
\begin{document}

\begin{frontmatter}

\title{Classicalization of Quantum Variables and Quantum-Classical Hybrids}
\author{T. Koide}
\address{ Instituto de F\'{\i}sica, Universidade Federal do Rio de Janeiro, C.P.
68528, 21941-972, Rio de Janeiro, Brazil }
\begin{abstract}
The extraction of classical degrees of freedom in quantum mechanics is studied in the 
stochastic variational method.
By using this classicalization, 
a hybrid model constructed from quantum and classical variables (quantum-classical hybrids) 
is derived. In this procedure, conservation laws such as energy are maintained, and 
Ehrenfest's theorem is still satisfied with modification.
The criterion for the applicability of quantum-classical hybrids is also discussed.
\end{abstract}


\end{frontmatter}

\section{Introduction}

Microscopic degrees of freedom obey the law of quantum mechanics (QM). On the other hand, 
classical mechanics (CM) is known to be an effective theory of QM and describes macroscopic physics precisely.
In other words, CM and QM are theories belonging to different dynamical hierarchies. 
Another example of such a hierarchy is observed in dynamics of fluids, 
where macroscopic behaviors of its constituent particles can be described 
by a coarse-grained dynamics called hydrodynamics.
However, particles which have much larger scales are suspended in a fluid, 
we need to solve coupled equations of the particles and the fluid \cite{push}.
This is an example of the appearance of a hybrid dynamics of different hierarchies.
Then it is interesting to ask whether we can construct an effective hybrid dynamics 
when quantum and classical degrees of freedom coexist.

In fact, there are several situations where such quantum-classical hybrids (QCH) seem to show up: 
quantum measurement \cite{sudar}, quantum-to-classical transition in early universe \cite{kiefer}, 
Einstein gravity interacting with quantum objects \cite{diosi}, Berry's phase \cite{zhang} and so on. 
Moreover, QCH has been studied to simplify complex numerical simulations 
in quantum chemistry \cite{bur}. 
See also Refs.\ \cite{hall1,elze} and the references therein.

There are already many models of QCH, but still no established theory.
The successful theories are constructed so as to combine 
quantum and classical systems consistently \cite{hall1,elze,buric}.
In this paper, we study this problem from a different perspective.
We first prepare a quantum system and derive a model of QCH by 
employing ``{\it classicalization}", which is a systematic method to replace  
quantum degrees of freedom with classical variables \cite{others}.


\section{Quantization of two-particle system}

Our classicalization is based on the hypothetical view that quantization can be formulated in the 
form of a variational principle by using the stochastic variational method (SVM), 
which was proposed by Yasue \cite{yasue} so as to reformulate Nelson's stochastic quantization \cite{nelson}.
As a pedagogical introduction of this approach, see Ref.\ \cite{koide-manual} and also Refs.\ \cite{zam,koide1,loff}. 
Before discussing QCH, we briefly introduce SVM as a quantization method.

Let us consider a non-relativistic two-particle system with the masses $m_1$ and $m_2$, and an interaction potential $V$.
Newton's equation of motion of this system is given by optimizing the action defined by $K-V$ with $K$ being a kinetic term.
Yasue showed that the Schr\"{o}dinger equation is derived by employing SVM to this classical action, 
replacing particle trajectories with stochastic variables. 

The stochastic trajectories of the two particles $\hat{\bf r}_1(t)$ and $\hat{\bf r}_2 (t)$ 
are characterized by the stochastic differential equations (i=1,2), 
\begin{equation}
\hspace*{-0.5cm} 
d\hat{\bf r}_i (t) = {\bf u}(\hat{\bf r}_1(t),\hat{\bf r}_2 (t),t)dt + \sqrt{2\nu_i} d{\bf W}_i(t)~~~~{ dt > 0}, \label{fsde} 
\end{equation}
where $dA(t) \equiv A(t+dt) - A(t)$, and ${\bf W}_1 (t)$ and ${\bf W}_2 (t)$ are independent Wiener processes.
The parameter $\nu_i$ characterizes the noise intensity.
When $\nu_i =0$, this coincides with the classical definition of particle velocity.
In other words, the origin of quantum fluctuation is attributed to the fluctuation of the particle trajectory in this formulation.

The unknown function ${\bf u}$ appearing here is determined by a variational principle. 
Let us consider a two particle classical system whose Lagrangian is given by 
${\displaystyle L = \sum_{i=1,2}m_i \dot{\bf r}^2_i/2 - V}$.
To quantize this classical system, we obtain the corresponding stochastic Lagrangian 
by substituting $\hat{\bf r}_i (t)$ as 
\begin{eqnarray}
\hspace*{-0.5cm}  L =
\sum_{i=1,2} \frac{m_i}{2} \frac{(D\hat{\bf r}_i (t))^2 + (\tilde{D}\hat{\bf r}_i(t))^2 }{2} 
- V(\hat{\bf r}_1(t),\hat{\bf r}_2(t)). \label{sl1} 
\end{eqnarray}
Here, two different time derivatives, $D$ and $\tilde{D}$, are introduced, because 
stochastic trajectories have zig-zag forms and the classical definition of velocity is not applicable. As was discussed by Nelson \cite{nelson}, 
the mean forward derivative $D$ and the mean backward derivative $\tilde{D}$ are defined by  
\begin{eqnarray}
\hspace*{-0.5cm}
D \hat{\bf r}_i (t) &=& \lim_{dt \rightarrow 0+} E \left[ \frac{\hat{\bf r}_i (t + dt) - \hat{\bf r}_i (t)}{dt} \Big| {\cal P}_t \right], \\
\hspace*{-0.5cm}
\tilde{D} \hat{\bf r}_i(t) &=& \lim_{dt \rightarrow 0-} E \left[ \frac{\hat{\bf r}_i (t + dt) - \hat{\bf r}_i (t)}{dt} \Big| {\cal F}_t \right],
\end{eqnarray}
respectively. Here $E[~~]$ denotes the expectation value. In particular, the above derivatives are 
defined as conditional averages, where ${\cal P}_t$ (${\cal F}_t$) indicates to fix 
$\hat{\bf r}(t')$ for $t'\le t$ $(t'\ge t)$.
Note that, by using Ito's lemma and the martingale of the Wiener process, these are calculated as  
\begin{eqnarray*}
\lefteqn{D f( \{ \hat{\bf r}(t) \} , t)} && \nonumber \\
&&\hspace*{-0.5cm}= \left[ \partial_t + \sum_{j=1,2} \left\{ {\bf u}_j ( \{ \hat{\bf r}(t) \} , t)\cdot \nabla_j + \nu_j \nabla^2_j \right\} \right] f( \{ \hat{\bf r}(t) \} , t), \\
\lefteqn{\tilde{D} f( \{ \hat{\bf r}(t) \} , t)} && \nonumber \\
&&\hspace*{-0.5cm} = \left[ \partial_t + \sum_{j=1,2} \left\{ \tilde{\bf u}_j ( \{ \hat{\bf r}(t) \} , t)\cdot \nabla_j - \nu_j \nabla^2_j \right\} \right] f( \{ \hat{\bf r}(t) \} , t), 
\end{eqnarray*}
where $\{ {\bf x} \} = ({\bf x}_1, {\bf x}_2)$ and 
$f$ is an arbitrary smooth function of the stochastic variable $\hat{\bf r}_i (t)$ and time $t$.
Note that $\tilde{\bf u}$ is defined by  
\begin{eqnarray*}
\tilde{\bf u}_i ( \{ {\bf x} \},t) \equiv {\bf u}_i (\{ {\bf x} \},t) - 2 \nu_i \nabla_i \ln \rho (\{ {\bf x} \},t),
\end{eqnarray*}
and this is called consistency condition.
Here $\rho ( \{ {\bf x} \},t)$ represents the two-particle probability density given by the solution of 
the equation of continuity,
\begin{equation}
\partial_t \rho (\{ {\bf x} \},t) = - \sum_{i=1,2} \nabla_i \cdot \{\rho (\{ {\bf x} \},t) {\bf v}_i(\{ {\bf x} \},t) \},
\label{fp}
\end{equation}
where  
\begin{eqnarray*}
&& {\bf v}_i  (\{ {\bf x} \},t) 
 \equiv \frac{{\bf u}_i  (\{ {\bf x} \},t) + \tilde{\bf u}_i  (\{ {\bf x} \},t)}{2}.
\end{eqnarray*}

Using these definitions, we can calculate the variation 
of the stochastic trajectories $\hat{\bf r}_i (t)$ following Ref.\ \cite{koide-manual}.
The result is known as stochastic Euler-Lagrange equation,
\begin{eqnarray}
\hspace*{-0.5cm}
 \left[ \tilde{D} \frac{\partial L}{\partial D \hat{\bf r}_i (t)} + D \frac{\partial L}{\partial \tilde{D} \hat{\bf r}_i (t)} 
- \frac{\partial L}{\partial \hat{\bf r}_i (t)} \right]^{\hat{\bf r}_1 (t) = {\bf x}_1 }
_{\hat{\bf r}_2 (t) = {\bf x}_2 } = 0 .\label{sel}
\end{eqnarray}
Note that $\hat{\bf r}_i (t)$ is replaced with a spatial parameter 
${\bf x}_i$ at the last of the calculation. 
It is because SVM requires that a stochastic action is optimized 
for any stochastic configuration of $\hat{\bf r}_i (t)$. 
Substituting the stochastic Lagrangian (\ref{sl1}) into this, we obtain 
\begin{eqnarray} 
&&\hspace*{-1.5cm} \left[ \partial_t   
+ \sum_{j=1,2} {\bf v}_j \cdot \nabla_j \right] {\bf v}_i   
= 2\nu_i \nabla_i \left[ \frac{1}{\sqrt{\rho}} \sum_{j=1,2} \nu_j \nabla^2_j 
\sqrt{\rho} \right] \nonumber \\
&&\hspace*{2cm} - \frac{1}{m_i} \nabla_i V({\bf x}_1,{\bf x}_2). \label{veq}
\end{eqnarray}
One can see that Eq.\ (\ref{veq}) is reduced to Newton's equation of motion in the vanishing limit of $\nu_i$.

Equations\ (\ref{fp}) and (\ref{veq}) are expressed in the form of the two-particle Schr\"{o}dinger equation, 
\begin{eqnarray*}
\hspace*{-0.8cm}
i \hbar \partial_t \psi (\{ {\bf x} \}, t)
= \left[ \sum_{i=1,2} \left\{ -\frac{\hbar^2}{2m_i} \nabla^2_i \right\} 
+ V(\{ {\bf x} \}) 
\right] \psi (\{ {\bf x} \}, t), \label{tpse}
\end{eqnarray*}
where the wave function is introduced by
\begin{eqnarray*}
\psi (\{ {\bf x} \}, t) = \sqrt{\rho(\{ {\bf x} \}, t) } e^{i \theta (\{ {\bf x} \}, t)},
\end{eqnarray*}
with the phase being defined by 
\begin{equation}
m_i {\bf v}_i (\{ {\bf x} \}, t) = \hbar \nabla_i \theta (\{ {\bf x} \}, t).\label{phase}
\end{equation} 
To obtain this result, we have already chosen 
\begin{equation}
\nu_i = \frac{\hbar}{2m_i}. \label{noise-intensity}
\end{equation}
That is, the noise intensity should be proportional to the inverse of the particle mass to 
reproduce QM.

The result of optimization depends on the definition of $d\hat{\bf r}_i(t)$.
If we assume the usual definitions of the classical velocity by 
taking the vanishing limit of $\nu_i\ (i=1,2)$ in Eqs.\ (\ref{fsde}) (as will be done in Eq.\ (\ref{app-qch}) only for $i=1$), 
Newton's equation of motion is obtained instead of the Schr\"{o}dinger equation. 
That is, SVM describes QM and CM in a unified way and this property is useful to discuss QCH.

The framework of SVM itself is more than the method of quantization and various applications are found in, for example, Ref.\ \cite{koide1}.

\section{QCH of two-particle system}

Let us derive a model of QCH under the assumption 
that the particle 1 is approximately described as a classical particle for a certain initial condition.
Then we need to {\it classicalize} the degree of freedom of the particle 1.
As we have seen, the origin of quantum fluctuation is attributed to the fluctuation of $\hat{\bf r}_i (t)$ induced by 
the Wiener process in SVM. 
Therefore, instead of Eq.\ (\ref{fsde}), we employ the following equation only for the particle 1 as the procedure of the classicalization, 
\begin{equation}
d\hat{\bf r}_1 (t) = {\bf v}_1 (\hat{\bf r}_1(t),\hat{\bf r}_2 (t),t) dt. \label{app-qch}
\end{equation}
Here $dt$ takes positive and negative values.
This is nothing but the classical definition of particle velocity.
This modification immediately affects the calculation of the mean derivatives as 
\begin{eqnarray*}
D \hat{\bf r}_1 (t) = \tilde{D} \hat{\bf r}_1 (t) = {\bf v}_1 (\hat{\bf r}_1(t),\hat{\bf r}_2 (t),t).
\end{eqnarray*}
On the other hand, the same definition (\ref{fsde}) is still used for $d\hat{\bf r}_2(t)$.

Remember that the noise intensity is proportional to the inverse of the mass of particles and thus 
the above approximation will be justified for heavier particles.  
Of course, the applicability of the classicalization depends also on initial conditions.
A more precise criterion of this applicability will be discussed later.

This modification of the definition drastically changes the result of the variation.
In fact, the stochastic Euler-Lagrange equation now becomes
\begin{eqnarray}
\left[ \tilde{D} \frac{\partial L}{\partial D \hat{\bf r}_i (t)} + D \frac{\partial L}{\partial \tilde{D} \hat{\bf r}_i (t)} 
- \frac{\partial L}{\partial \hat{\bf r}_i (t)} \right]^{\hat{\bf r}_1 (t) = {\bf f}_1({\bf x}_2,t)}
_{\hat{\bf r}_2 (t) = {\bf x}_2 } = 0 .\label{sel-qch}
\end{eqnarray}
Here, the substitution of $\hat{\bf r}_1 (t)$ is changed from ${\bf x}_1$ to ${\bf f}_1({\bf x}_2,t)$.
As mentioned already, SVM requires that the action is optimized for any configuration of 
stochastic trajectories and thus $\hat{\bf r}_2 (t)$ is replaced by the spatial parameter ${\bf x}_2$. 
However, as is seen from Eq.\ (\ref{app-qch}), $\hat{\bf r}_1 (t)$ fluctuates because of 
the $\hat{\bf r}_2(t)$ dependence in ${\bf v}_1$. 
Once ${\bf x}_2$ is substituted into $\hat{\bf r}_2 (t)$, $\hat{\bf r}_1 (t)$ does not fluctuate any more 
and hence cannot be replaced simply by ${\bf x}_1$.
In this case, the ``velocity" on the right hand side of Eq.\ (\ref{app-qch}) 
becomes a function not only of $t$ but also of ${\bf x}_2$. 
Therefore we introduce a new quantity defined by the solution of the following equation,
\begin{equation}
\partial_t {\bf f}_1 ({\bf x}_2,t) = {\bf v}_1 ({\bf f}_1 ({\bf x}_2,t),{\bf x}_2,t). \label{pse-tra}
\end{equation}
We call ${\bf f}_1 ({\bf x}_2,t)$ quasi trajectory function, which  
is substituted into $\hat{\bf r}_1 (t)$ in Eq.\ (\ref{sel-qch}).

The variations of Eq.\ (\ref{sel-qch}) by $\hat{\bf r}_1(t)$ and $\hat{\bf r}_2(t)$, respectively, lead to  
\begin{eqnarray}
&& \hspace*{-2cm} 
\left[ \frac{d}{dt} + \frac{\hbar}{m_2} \nabla_2 {\rm Im} (\ln \Psi_2 ({\bf f}_1({\bf x}_2,t),{\bf x}_2,t)) \cdot \nabla_2
\right]
 {\bf v}_1 ({\bf f}_1({\bf x}_2,t),{\bf x}_2,t)  \nonumber \\
&& = - \frac{1}{m_1} \nabla_1 V({\bf f}_1({\bf x}_2,t),{\bf x}_2), \label{v1eq} 
\end{eqnarray}
and
\begin{eqnarray}
\hspace*{-1cm} 
 i\hbar \frac{d}{dt} \Psi_2 ({\bf f}_1({\bf x}_2,t),{\bf x}_2,t) 
= \hat{H}_2 (t) \Psi_2 ({\bf f}_1({\bf x}_2,t),{\bf x}_2,t).  \label{psi22}
\end{eqnarray}
Here, note that the total time derivative is defined by 
\begin{eqnarray}
\frac{d}{dt} \equiv \partial_t + {\bf v}_1 ({\bf f}_1({\bf x}_2,t),{\bf x}_2,t)  \cdot \nabla_1, \label{ddt}
\end{eqnarray}
and the time evolution operator is given by 
\begin{eqnarray}
&& \hspace*{-1cm} \hat{H}_2 (t) = -\frac{m_1}{2} {\bf v}^2_1 ({\bf f}_1({\bf x}_2,t),{\bf x}_2,t) - \frac{\hbar^2}{2m_2} \nabla^2_2 
\nonumber \\
&& + V({\bf f}_1({\bf x}_2,t),{\bf x}_2). \label{ham2}
\end{eqnarray}
The wave function for the particle 2 is denoted by $\Psi_2 ({\bf f}_1({\bf x}_2,t),{\bf x}_2,t)$ which gives the probability density of the particle 2 by 
$|\Psi_2|^2 = \rho_2({\bf x}_2,t)$. 
The phase of $\Psi_2$ is defined in a similar way to Eq.\ (\ref{phase}).

In short, our QCH model consists of Eqs.\ (\ref{pse-tra}), (\ref{v1eq}) and (\ref{psi22}).
Note that Eq.\ (\ref{psi22}) is expressed by the total time derivative $d/dt$, but if we 
re-express it by $\partial_t$ using Eq.\ (\ref{ddt}), the sign of the first term in Eq.\ (\ref{ham2}) becomes positive.

\section{Properties of QCH}

\subsection{No interaction limit}

If there is no interaction and no initial correlation between the particles 1 and 2, 
${\bf v}_1$ is independent of ${\bf x}_2$ from Eq.\ (\ref{v1eq}), and hence 
the quasi trajectory function coincides with the classical trajectory ${\bf r}_1 (t)$, because
$d {\bf f}_1 (t)/dt = d {\bf r}_1 (t)/dt = {\bf v}_1 ({\bf r}_1 (t),t)$.
Then Eq.\ (\ref{v1eq}) is reduced to Newton's equation of motion.
On the other hand, the first term in Eq.\ (\ref{ham2}) becomes a function only of a time and 
such a term can be absorbed into the phase of the wave function. 
Then Eq.\ (\ref{psi22}) is reduced to 
the usual Schr\"{o}dinger equation in this limit.

Note that the difference between ${\bf f}_1$ and $\hat{\bf r}_1$ plays an important role 
in Ehrenfest's theorem.

\subsection{Energy conservation}

Usually, the definition of conserved quantities is not trivial when quantum and classical variables coexist. 
The advantage of our approach is that any conserved quantity is defined directly from 
the invariance of the stochastic action through the stochastic Noether theorem \cite{misawa}. 
The energy, for example, obtained from Eq.\ (\ref{sl1}) is given by  
\begin{eqnarray}
&& \hspace*{-1.5cm} E 
= 
\left\langle 
\sum_{i=1}^2 \frac{m_i}{2}{\bf v}^2_i ({\bf f}_1 ({\bf x}_2,t),{\bf x}_2 ,t ) 
+ V({\bf f}_1({\bf x}_2,t),{\bf x}_2) \right. \nonumber \\
&& \hspace*{-0.5cm} \left. + \frac{\hbar^2}{8m_2} (\nabla_2 \ln |\Psi_2({\bf f}_1 ({\bf x}_2,t),{\bf x}_2,t)|^2)^2
\right\rangle_2,
\end{eqnarray}
where 
\begin{eqnarray*}
\langle A \rangle_2 \equiv \int d^3 {\bf x}_2 \Psi_2^* ({\bf f}_1({\bf x}_2,t),{\bf x}_2,t) A \Psi_2 ({\bf f}_1({\bf x}_2,t),{\bf x}_2,t),
\end{eqnarray*} 
for an arbitrary function $A$. 
One can confirm that this quantity is conserved by the direct substitution of our equations. 
The conservations of momentum and angular momentum also can be obtain in a similar way.

\begin{figure}[t]
\begin{center}
\includegraphics[scale=0.18]{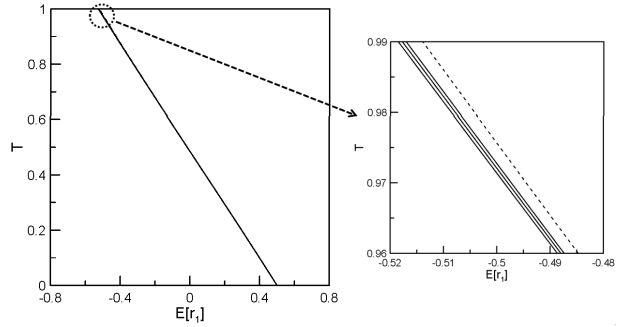}
\caption{The time evolution of $E[ \hat{r}_1 (t)]$. The four lines lie on top of each other. 
The extended figure is shown in the right panel. The solid and dashed lines represent the results of QCH and QM, respectively. 
The solid lines correspond to $\alpha = 0.5,1$ and $2$ from the most left line. }
\label{fig:cla}
\end{center}
\end{figure}

\subsection{Ehrenfest's theorem}

Ehrenfest's theorem is satisfied for our QCH model, but in a modified way.
The expectation values of $\hat{\bf r}_1(t)$, ${\bf v}_1$, ${\bf x}_2$ and ${\bf v}_2$ of our model 
are given by
\begin{eqnarray}
&& \hspace*{-1cm}\frac{d}{dt} E[ \hat{\bf r}_1(t) ] \nonumber \\
&& = \langle {\bf v}_1({\bf f}_1({\bf x}_2,t),{\bf x}_2,t) \rangle_2, \label{exp-r1}\\
&& \hspace*{-1cm} \frac{d}{dt} \langle {\bf v}_1({\bf f}_1({\bf x}_2,t),{\bf x}_2,t) \rangle_2 \nonumber \\
&& = - \frac{1}{m_1} \langle \nabla_1 V({\bf f}_1({\bf x}_2,t),{\bf x}_2) \rangle_2, \qquad 
\end{eqnarray}
for the particle 1, and 
\begin{eqnarray}
\frac{d}{dt} \langle {\bf x}_2 \rangle_2 &=& \frac{1}{m_2} \langle \hat{\bf p}_2 \rangle_2, \\
\frac{d}{dt} \langle \hat{\bf p}_2 \rangle_2 &=& -\langle \nabla_2 V({\bf f}_1({\bf x}_2,t),{\bf x}_2) \rangle_2,
\end{eqnarray} 
for the particle 2, respectively.
Here $\hat{\bf p}_2 = -i\hbar \nabla_2$ and we used 
\begin{eqnarray*}
\hspace*{-0.5cm}
\left[ \frac{d}{dt}, \hat{\bf p}_2 \right] = [\partial_t + {\bf v}_1 \cdot \nabla_1, \hat{\bf p}_2] 
= i\hbar \sum_{j=1}^3 (\nabla_2 {\bf v}^j_1) \nabla^j_1.
\end{eqnarray*}
Note that Eq.\ (\ref{exp-r1}) is obtained from Eq.\ (\ref{app-qch}) and corresponds to the (mean) 
trajectory of the particle 1.

\begin{figure}[t]
\begin{center}
\includegraphics[scale=0.18]{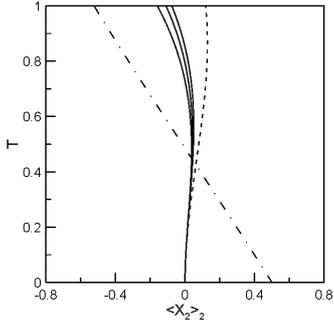}
\caption{The time evolution of $\langle x_2 \rangle_2$. The solid and dashed lines represent the results of QCH and QM, respectively. 
The solid lines correspond to $\alpha = 0.5,\ 1$ and $2$ from the most left line. For comparison, the dot-dashed line represents the trajectory of the particle 1 
shown in Fig. \ref{fig:cla}.}
\label{fig:qua}
\end{center}
\end{figure}

One can see that the above four equations correspond to Ehrenfest's theorem in QM when 
${\bf f}_1({\bf x}_2,t)$ coincides with the particle trajectory  $\hat{\bf r}_1 (t)$.
To study the difference of these quantities, 
let us consider a harmonic potential 
\begin{eqnarray*}
V({\bf x}_1,{\bf x}_2) = \frac{K}{2} ({\bf x}_1 - {\bf x}_2)^2, 
\end{eqnarray*}
where the behaviors of 
the expectation values $\langle \hat{\bf x}_i \rangle$ and $\langle \hat{\bf p}_i \rangle$ in QM  
exactly agree with Newton's equations of motion.
That is, the evolutions are independent of the detailed form of the wave function by fixing 
the initial values of $\langle \hat{\bf x}_i \rangle$ and $\langle \hat{\bf p}_i \rangle$.
However, it is not the case for our QCH model and the behaviors depend on the forms of the wave function around 
$\langle \hat{\bf x}_i \rangle$ and $\langle \hat{\bf p}_i \rangle$.

We consider a one-dimensional system with $m_1 = 5 m_2$ and $K = \hbar^2  /(2m_2 \lambda^4)$ with 
$\lambda$ being a parameter with a spatial dimension.
Positions are divided by $\lambda$ and adimensional, and a dimensionless time is given by $T \equiv \hbar t /(2m_2 \lambda^2)$. 
The initial conditions are given by $\hat{r}_1(0) = f_1({x}_2,0) = 0.5$, 
${v}_1 (f_1({x}_2,0),x_2,0) =  - \hbar/(2m_2 \lambda)$ and 
\begin{eqnarray}
\Psi_2(f_1({x}_2,0),x_2,0) = \left( \frac{2\alpha}{\pi} \right)^{1/4}e^{-\alpha x^2_2}. \label{psi2}
\end{eqnarray}
Here $\alpha$ is the width of $\Psi_2$ and we consider three cases, 
$\alpha = 0.5$, $1$ and $2$.

\begin{figure}[t]
\begin{center}
\includegraphics[scale=0.18]{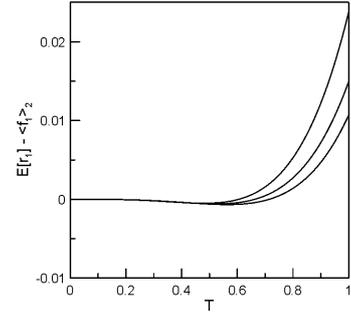}
\caption{The difference of the two trajectories $E[\hat{r}_1 (t)] - \langle x_2 \rangle_2$. The solid lines represent $\alpha = 0.5,\ 1$ and $2$ from the top line.}
\label{fig:ratio}
\end{center}
\end{figure}

In Fig.\ \ref{fig:cla}, the time evolution of the particle trajectory $E[ \hat{r}_1(t) ]$ is plotted. The four lines lie on top of each other 
and the extended figure is shown on the right panel. 
The solid and dashed lines represent the results of QCH and QM, respectively. 
The solid lines correspond to $\alpha = 0.5$, $1$ and $2$ from the most left line.
We can see that the QCH model becomes better approximation of QM as $\alpha$ increases.

The difference between QCH and QM is more remarkable for $\langle x_2 \rangle_2$ shown in Fig. \ref{fig:qua}. 
Again, the solid and dashed lines represent the results of QCH and QM, and the solid lines correspond to $\alpha = 0.5$, $1$ and $2$ from the most left line. 
For comparison, the dot-dashed line represents the same trajectory of the particle 1 
shown in Fig. \ref{fig:cla}. The particles 1 and 2 cross around $T \sim 0.4$.
Again, the deviations from QM become smaller as $\alpha$ increases.

This deviation is related to the limited applicability of QCH, where 
a wave function with a finite width 
is approximately replaced by a point-like particle for the particle 1. 
This approximation is justified as the distance between the two particles 1 and 2 enlarges.
Increases the value of $\alpha$, decreases the overlap between the two particles. 
It is effectively equivalent to separate off the two particles.
This is why our QCH model becomes a better approximation of QM as $\alpha$ increases.

This can be seen even from the behavior of the quasi trajectory defined by 
$\langle {\bf f}_1 ({\bf x}_2,t) \rangle_2$.
Its differential equation is given by 
\begin{eqnarray}
&& \frac{d}{dt} \langle {\bf f}_1({\bf x}_2,t) \rangle_2 = \langle {\bf v}_1 ({\bf f}_1({\bf x}_2,t), {\bf x}_2,t) \rangle_2 
\nonumber \\
&& +\frac{i}{\hbar} \langle [\hat{H}_2 (t), {\bf f}_1({\bf x}_2,t)] \rangle_2.
\end{eqnarray}
For the QCH model to be a reasonable approximation of QM, Ehrenfest's theorem should be satisfied and thus the 
quasi trajectory should coincide with the trajectory. 
Comparing to Eq.\ (\ref{exp-r1}), 
the quasi trajectory can be identified with the trajectory $E[\hat{r}_1 (t)]$ itself 
when the contribution from the second term on the right hand side of the above equation is sufficiently small. 
In fact, as we have seen, this term vanishes when there is no interaction between the particles 1 and 2.

In Fig.\ \ref{fig:ratio}, the difference of the trajectories, $E[\hat{r}_1 (t)] - \langle f_1(x_2,t) \rangle_2$, 
is shown. 
The difference starts to grow up around $T\sim 0.4$ where the two particles intersect 
and the QCH model is not a reasonable approximation. 
We further observe that, as $\alpha$ increases, the difference decreases.
These observations confirm our argument of the limited applicability.
As was discussed, a wave function with a finite width is replaced by a 
point-like particle in the approximation of QCH. When the classical and quantum parts are well-separated, 
such a replacement will not give rise to a significant problem. 
In the present case, however, the distance between the two particles is shorter than the width of the wave function. Thus the difference between the two trajectories 
continues increasing during the time evolution. This increase will stop when these particles move away from each other.
Note that this difference can be used to characterize the applicability of the QCH approach.

In the above simulations, we fixed the mass ratio by $m_1/m_2=5$. 
It is confirmed numerically that the deviation from QM decreases even by increasing this ratio.

\begin{figure}[t]
\begin{center}
\includegraphics[scale=0.16]{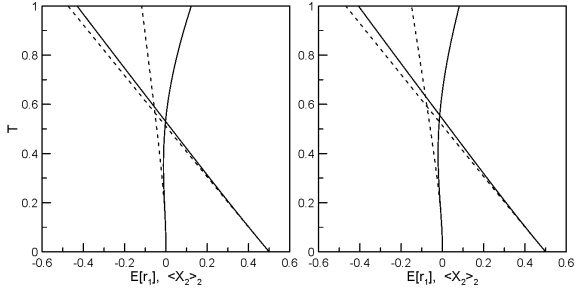}
\caption{The time evolutions of $E[\hat{r}_1(t)]$ (starting from $0.5$) and $\langle x_2 \rangle_2$ (starting from $0$). 
The solid and dashed lines represent the results of QCH and QM, respectively. 
The left panel is the results for $\alpha = 0.5$ and the right panel for $\alpha = 1$, respectively.
}
\label{fig:another-cla-qua-tra}
\end{center}
\end{figure}

\subsection{Application to a repulsive potential}

The above properties might be the consequence of a special property attributed to the harmonic potential where 
trajectories in CM and QM exactly coincide.
Thus it is important to confirm the behaviors of our QCH model in other potentials.

Let us consider the following repulsive interaction, 
\begin{eqnarray*}
V (x_1,x_2)= \frac{\hbar^2}{2m_2 \lambda^2} e^{-(x_1 - x_2)^2}.
\end{eqnarray*}
As for the initial conditions to solve QCH, we employ the same values as the previous harmonic potential calculation, which are given around Eq.\ (\ref{psi2}).
To solve the two-particle Schr\"{o}dinger equation with this potential, on the other hand, we adapt 
\begin{eqnarray*}
\Psi(x_1,x_2,0) = \left( \frac{8\alpha}{\pi^2} \right)^{1/4} e^{-i 5 x_1 - 2 (x_1 - 1/2)^2}
e^{- \alpha x_2^2} \nonumber,
\end{eqnarray*}
where $\alpha$ is the width of the initial wave function, which already appears in Eq.\ (\ref{psi2}).
(Note that, in the previous harmonic potential calculation, the trajectories $\langle \hat{x}_i \rangle$ in QM are exactly 
the same as the corresponding classical ones and we did not need to specify the two-particle initial wave function in detail.)

The time evolutions of $E[\hat{r}_1 (t)]$ and $\langle x_2 \rangle_2$ are shown in Fig.\ \ref{fig:another-cla-qua-tra}; 
the former is the almost straight line starting from $0.5$ and the latter the curved line from $0$.  
The left and right panels represent the results of $\alpha = 0.5$ and $1$, respectively. 
The solid lines denote the results of QCH and the dashed lines are $\langle \hat{x}_i \rangle$ 
obtained from the two-particle Schr\"{o}dinger equation.
The deviation from the QM result is large for $\alpha = 0.5$, and 
as was discussed in the previous section, this can be explained by the fact that 
QCH looses its validity as the increase of the overlap between the wave functions of the two particles. 

\begin{figure}[t]
\begin{center}
\includegraphics[scale=0.18]{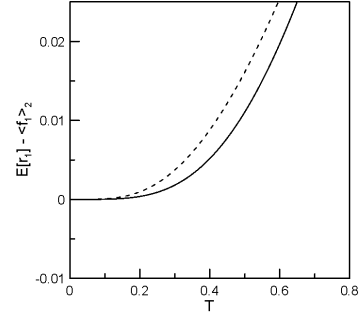}
\caption{The difference of the two trajectories $E[\hat{r}_1 (t)] - \langle x_2 \rangle_2$ plotted.
The dashed and the solid lines represent the results of $\alpha=0.5$ and $1$, respectively.
}
\label{fig:ratio-another2}
\end{center}
\end{figure}

This is more visible from the difference of the two trajectories, $E[\hat{r}_1 (t)] - \langle f_1(x_2,t) \rangle_2$.
The result is shown in Fig.\ \ref{fig:ratio-another2}, where the dashed and solid lines represent the results of $\alpha = 0.5$ and $1$, respectively. One can see that the dashed line deviates from $0$ more rapidly.

These behaviors are consistent with what we observed in the harmonic potential calculation and our QCH model is 
expected to be reliable even for general potentials.

\section{QCH of $N_c + N_q$-particle system}

Our result can be extended to a many particle system which 
consists of $N_c$ classical $\{ \hat{\bf r} \}$ and $N_q$ quantum $\{ \hat{\bf q} \}$ variables. 
The stochastic Lagrangian is expressed as 
\begin{eqnarray}
L &=& \sum_{a=1}^{N_c} \frac{m_a}{2} \frac{(D\hat{\bf r}_a)^2 + (\tilde{D}\hat{\bf r}_a)^2 }{2} \nonumber \\
&& \hspace*{-1cm}+ \sum_{\alpha=1}^{N_q} \frac{m_\alpha}{2} \frac{(D\hat{\bf q}_\alpha)^2 + (\tilde{D}\hat{\bf q}_\alpha)^2 }{2}
- V(\{{\bf r} \},\{\hat{\bf q}\}),
\end{eqnarray}
where $V =V(\{\hat{\bf r} \},\{\hat{\bf q}\})$ is an interaction potential. 
The stochastic variations lead to 
\begin{eqnarray}
&&\hspace*{-1cm} \left[ \frac{d}{dt}  + \sum_{\beta=1}^{N_q} \frac{\hbar}{m_\beta} \nabla_\beta {\rm Im} (\ln \Psi_q) \cdot \nabla_\beta  \right] {\bf v}_a 
= 
- \frac{1}{m_a} \nabla_a V, \qquad \\
&&  i \hbar \frac{d}{dt}\Psi_q (\{ {\bf f} \}, \{ {\bf x} \}, t) 
= \hat{H}_q (t) \Psi_q (\{ {\bf f} \}, \{ {\bf x} \}, t),
\end{eqnarray}
where $d/dt = \partial_t + {\displaystyle \sum_{a=1}^{N_c}} {\bf v}_a \cdot\nabla_a$, and 
\begin{eqnarray}
&&\hspace*{-1cm} \partial_t {\bf f}_a (\{ {\bf x} \},t) = {\bf v}_a (\{ {\bf f}  \}, \{ {\bf x} \},t), \\
&&\hspace*{-1cm} \hat{H}_q (t) =  - \sum_{a=1}^{N_c} \frac{m_a}{2} {\bf v}^2_a 
- \sum_{\alpha=1}^{N_q} \frac{\hbar^2}{2m_\alpha} \nabla_\alpha^2 + V .
\end{eqnarray}

\section{Concluding Remarks}

In this paper, we formulated classicalization of quantum variables based on SVM and 
derived a QCH model. 
This model conserves any quantity associated with the invariant transforms of the action 
and satisfies the extended Ehrenfest theorem.
Our QCH model is a good approximation of QM when there exists a large mass difference among particles 
and no significant overlap of quantum states. 
That is, we should use initial conditions where classical 
and quantum particles are well-separated each other in the application of QCH.
The failure of QCH can be estimated by calculating the difference 
between the quasi trajectory and the trajectory.
This quantity may play a role of an order parameter of the quantum-to-classical transition.

There are already many proposals for QCH, but most of the 
models cannot preserve important laws of physics and/or 
consistency requirements such as the energy conservation, the positivity of probability, Ehrenfest's theorem and so on \cite{elze}. 
So far there are three models which are considered to be 
reliable \cite{hall1,elze,buric}. See also an example of the QCH approach in Ref.\ \cite{ELF2}.
In such models, there is no concept of the quasi trajectory and thus ours is different from these successful models.
We do not find any inconsistency in our model so far, 
but the relation to these models should be investigated carefully.
For example, a QCH state in such models is characterized by the combination of the Hilbert space of QM 
and the phase space of CM, while it is not clear how the phase space is introduced in our model. 
The advantage of our approach is that conserved quantities are defined 
from the stochastic Noether theorem straightforwardly.

SVM is applicable to the field quantization \cite{koidef} and thus 
a QCH model of a field system will be constructed. 
This is left as a future task.

The author acknowledges H.-T.Elze, T. Kodama, K. Tsushima and participants of DICE2014 
for fruitful discussions and comments.
This work is financially supported by CNPq.

\appendix

\end{document}